\documentclass[fleqn,twoside]{article}
\usepackage{espcrc2,epsfig}

\usepackage{graphicx}
\usepackage[figuresright]{rotating}

\newcommand{\AmS}{{\protect\the\textfont2
  A\kern-.1667em\lower.5ex\hbox{M}\kern-.125emS}}

\title{Particle Acceleration at shocks: some modern aspects of an old problem}

\author{P. Blasi\address[INAF]{INAF/Osservatorio Astrofisico di Arcetri\\
    Largo E. Fermi, 5 - Firenze (Italy)}}
       
\begin{document}

\begin{abstract}
The acceleration of charged particles at astrophysical collisionless 
shock waves is one of the best studied processes for the energization 
of particles to ultrarelativistic energies, required by multifrequency 
observations in a variety of astrophysical situations. In this paper
we discuss some work aimed at describing one of the main progresses
made in the theory of shock acceleration, namely the introduction of the
non-linear backreaction of the accelerated particles onto the shocked
fluid. The implications for the investigation of the origin  of ultra 
high energy cosmic rays will be discussed.
\vspace{1pc}
\end{abstract}

\maketitle

\section{Introduction}

Suprathermal charged particles scattering back and forth across the 
surface of a shock wave gain energy. The concept of stochastic 
energization due to randomly moving inhomogeneities was first 
proposed by Fermi \cite{fermi}. In that original version, the 
acceleration process is easily shown to be efficient only at the
second order in the parameter $V$, the average speed of the irregularities 
in the structure of the magnetic field, in units of the speed of
light. For non-relativistic motion, $V\ll 1$, the mechanism is 
not very attractive. The generalization of this idea to the case of 
a shock wave was first proposed in \cite{als77,krimsky77,bell78,bo78} 
and is nicely summarized in several recent reviews 
\cite{be87,drury83,bk88,je91,md01}, where the 
efficiency of the process was found to be now at the first order
in $V$. Since these pioneering papers the process of particle
acceleration at shock waves has been investigated in many aspects
and is now believed to be at work in a variety of astrophysical 
environments. In fact we do observe shocks everywhere, from the 
solar system to the interplanetary medium, from the supernovae 
environments to the formation of the large scale structure
of the universe. All these are therefore sites of both heating 
of the medium crossing the shock surface and generation of 
suprathermal particles. The two phenomena are most likely 
different aspects of the same process, also responsible for the 
formation of the collisionless shock itself. One of the major
developments in the theory of particle acceleration at astrophysical
shock waves has consisted of removing the assumption of {\it test 
particle}, namely the assumption that the accelerated particles
could not affect the dynamics of the shocked fluid 
\footnote{In the presentation at CRIS 2004, 
particle acceleration at relativistic shocks was also discussed at 
length. That part is not included here, due to the limited space 
available.}. Two approaches
have been proposed to treat this intrinsically non-linear problem:
the {\it two fluid models} \cite{dr_v80,dr_v81,dr_ax_su82,ax_l_mk82,ddv94} 
and the {\it kinetic models} \cite{malkov1,malkov2,blasi02,blasi2004}, 
while numerous attempts to simulate numerically the process of particle 
acceleration have also been made 
\cite{je91,bell87,elli90,ebj95,ebj96,kj97,jones02}.
The two fluid models treat the accelerated particles as a separate fluid,
contributing a pressure and energy density which enter the usual 
conservation laws at the shock surface. By construction, these
models do not provide information about the spectrum of the 
accelerated particles, while correctly describing the detailed dynamics
of the fluids involved. The kinetic models on the other hand 
have a potential predictive power in terms of both dynamics and 
spectral shape of the accelerated particles. 

All these considerations hold in principle for all shocks but in
practice most of the work has been done for the case of newtonian
shock waves (however see \cite{ell_NL} for an extension to relativistic
shocks). Astrophysical studies have 
shown that there are plenty of examples in Nature of fluids moving at 
relativistic speeds, and generating shock waves. The generalization of the 
process of particle acceleration to the relativistic case represents
in our opinion the second major development of the theory (Baring, these
proceedings).

In this paper, we will not present a review of all the current efforts
in the investigation of shock acceleration. We will rather concentrate
our attention upon some recent work in the direction of 
accounting for the non-linear backreaction of the accelerated particles.

\section{The non-linear backreaction: breaking the test particle
approximation}

The original theory of particle acceleration was based on the 
assumption that the accelerated particles represent a {\it passive}
fluid, with no dynamical backreaction on the background plasmas involved. 
Within the context of this approximation, several independent approaches
\cite{bell78,be87} give the spectrum of the accelerated particles
in the form of a power law in momentum $f(p)=f_0 (p/p_0)^{-\alpha}$,
where the slope $\alpha$ is related in a unique way to the Mach
number $M$ of the upstream fluid as seen in the shock frame, through 
the expression $\alpha = 4M^2/(M^2-1)$ (here we asumed that the adiabatic 
index of the background gas is $\gamma_g=5/3$). This result is easily shown 
by using the diffusion-convection equation in one dimension for a stationary
situation (namely $\partial f/\partial t = 0$): 
$$
\frac{\partial}{\partial x}
\left[ D  \frac{\partial}{\partial x} f(x,p) \right] - 
u  \frac{\partial f (x,p)}{\partial x} + 
$$
\begin{equation}
+ \frac{1}{3} \frac{d u}{d x}~p~\frac{\partial f(x,p)}{\partial p} + 
Q(x,p) = 0,
\label{eq:trans}
\end{equation}
where $D$ is the diffusion coefficient, $f(x,p)$ is the distribution 
function of accelerated particles in phase space and $Q$ is the 
injection function, which we will assume to be a Dirac delta function
at the shock surface $x=0$ in the downstream fluid ($x=0^+$). The function $f$ 
is normalized in such a way that the total number of accelerated particles
is given by $\int_{p_{inj}}^{p_{max}} dp 4\pi p^2 f(p)$. 

As a first step, we integrate eq. \ref{eq:trans} around $x=0$, from $x=0^-$ to
$x=0^+$, which we denote as points ``1'' and ``2'' respectively, so that we get
\begin{equation}
\left[ D \frac{\partial f}{\partial x}\right]_2 -
\left[ D \frac{\partial f}{\partial x}\right]_1 +
\frac{1}{3} p \frac{d f_0}{d p} (u_2 - u_1) + Q_0(p)= 0,
\end{equation}
where $u_1$ ($u_2$) is the fluid speed immediately upstream (downstream) 
of the shock and $f_0$ is the particle distribution function at the shock
location. By requiring that the distribution function downstream is 
independent of the spatial coordinate (homogeneity), we obtain
$\left[ D \frac{\partial f}{\partial x}\right]_2=0$, so that the boundary 
condition at the shock can be rewritten as
\begin{equation}
\left[ D \frac{\partial f}{\partial x}\right]_1 =
\frac{1}{3} p \frac{d f_0}{d p} (u_2 - u_1) + Q_0(p).
\label{eq:boundaryshock}
\end{equation}
We can now perform the integration of eq. (\ref{eq:trans}) from $x=-\infty$ to
$x=0^-$ (point ``1''), in order to take into account the boundary condition at 
upstream infinity. Using eq. (\ref{eq:boundaryshock}) we obtain
\begin{equation}
\frac{1}{3} p \frac{d f_0}{d p} (u_2 - u_1) - u_1 f_0 + Q_0(p) = 0.
\label{eq:step}
\end{equation}
The solution of this equation for $f_0$ has the form of a power law $f_0 
\propto p^{-\alpha}$ with slope $\alpha=3 u_1/(u_1-u_2)=3r/(r-1)$, where 
we introduced the compression factor $r=u_1/u_2$ at the shock. For a 
strong shock $r\to 4$ and we find the well known asymptotic spectrum 
$f_0\to p^{-4}$, or $N(E)\propto E^{-2}$ in terms of energy (here again
we assumed that the adiabatic index of the background gas is $\gamma_g=
5/3$. 

Why should we expect this simple result to be affected by the assumption
of test particles? There are three physical arguments that may serve
as plausibility arguments to investigate the effects of possible 
backreactions: 1) the spectrum $E^{-2}$ is logarithmically divergent
in its energy content, so that even choosing a maximum momentum, it is 
possible that the energy density in the form of accelerated particles 
becomes comparable with the kinetic pressure, making the assumption of 
test particles untenable;
2) if the non thermal pressure becomes appreciable, the effective adiabatic
index can get closer to $4/3$ rather than $5/3$, making the shock more 
compressive and the spectrum of accelerated particles even more divergent;
3) more divergent spectra imply larger fluxes of escaping particles at
the maximum momentum, which make the shock radiative-like, again implying
a larger compression and flatter spectra. 

All the three issues raised here point toward the direction of making the
backreaction more severe rather than alleviating its effect, therefore a 
run-away reaction seems likely, which drives the shock toward a strongly 
non-linear cosmic ray modified configuration (here the term {\it cosmic rays} 
is used in a general way to indicate the accelerated particles).  

We can describe the expected effects on the basis of the following simple 
argument: if, as is usually the case, the diffusion coefficient increases 
with the momentum of the particles, we can expect that particles with 
larger momenta will diffuse farther from the shock surface in the upstream
section of the gas. At large distances from the shock, only the high energy
particles will be present, while lower energy particles will populate the 
regions closer to the shock surface. There is some critical distance which 
corresponds to the typical diffusion length of the particles with the 
maximum momentum achievable, $p_{max}$. At this distance, the pressure of
the cosmic rays is basically zero and the fluid is unperturbed. On the 
other hand, moving inward, toward the shock, an increasing number of 
accelerated particles is present, and their pressure contributes to the
local pressure budget by slowing down the fluid (in the shock frame). 
This effect causes the fluid speed upstream to be space-dependent, and
decreasing while approaching the shock surface. The region of slow 
decrease of the fluid velocity is usually called the {\it precursor}. 
The shock, which may now be substantially weakened by the effect of 
the accelerated particles, is usually called {\it subshock}. 

It is useful to introduce the two quantities $R_{sub}=u_1/u_2$ and 
$R_{tot}=u_0/u_2$, which are respectively the compression factor at the 
gas subshock and the total compression factor between upstream infinity 
and downstream. Here $u_0$, $u_1$ and $u_2$ are the fluid speeds at
upstream infinity, upstream of the subshock and downstream respectively.
The two compression factors would be equal in the test particle approximation.
For a modified shock, $R_{tot}$ can attain values much larger than
$R_{sub}$ and more in general, much larger than $4$, which is the maximum value
achievable for an ordinary strong non-relativistic shock. 

The shape of the particle spectrum is still determined by some jump in the
velocity field, but this quantity is now local: at low energies, the 
compression felt by the particles is $\sim R_{sub}$, while at $p\sim p_{max}$
the effective compression is $\sim R_{tot}$. It follows that, since 
$R_{sub}<R_{tot}$, the spectrum at low energies is steeper than that 
at higher energies: the overall spectrum at cosmic ray modified shocks
is therefore expected to have a concave shape. 

In the following we will describe the effects of the particle backreaction
following the kinetic semi-analytical approach proposed in 
\cite{blasi02,blasi2004}, and we will use the most general formalism, which 
includes the possible presence of seed pre-accelerated particles in the
environment in which the shock propagates. We repeat here the steps 
illustrated above for the linear case.
Integrating again eq. \ref{eq:trans} around $x=0$, from $x=0^-$ to
$x=0^+$, we get Eq. \ref{eq:boundaryshock}, after invoking the homogeneity
of the particle distribution downstream.

Performing now the integration of eq. \ref{eq:trans} from $x=-\infty$ to
$x=0^-$ we obtain
$$
\frac{1}{3} p \frac{d f_0}{d p} (u_2 - u_1) - u_1 f_0 + u_0 f_\infty + Q_0(p)+
$$
\begin{equation}
\int_{-\infty}^{0^-} dx f \frac{d u}{d x} + \frac{1}{3}\int_{-\infty}^{0^-} 
dx  \frac{d u}{d x} p \frac{\partial f}{\partial p} = 0.
\label{eq:stepnl}
\end{equation}
Here $f_\infty$ represents the distribution of seed pre-accelerated
particles possibly present at upstream infinity.

We can now introduce the quantity $u_p$ defined as
\begin{equation}
u_p = u_1 - \frac{1}{f_0} \int_{-\infty}^{0^-} dx \frac{d u}{d x} f(x,p),
\label{eq:up}
\end{equation}
whose physical meaning is instrumental to understand the nonlinear backreaction
of the accelerated particles. The function $u_p$ is the average fluid 
velocity experienced by particles with momentum $p$ while diffusing 
upstream away from the shock surface. In other words, the effect of the 
average is that, instead of a constant speed $u_1$ upstream, a particle 
with momentum $p$ experiences a spatially variable speed, due to the 
pressure of the accelerated particles that is slowing down the fluid. Since 
the diffusion coefficient is in general $p$-dependent, particles with 
different energies {\it feel} a different compression coefficient, higher 
at higher energies if, as expected, the diffusion coefficient is an 
increasing function of momentum.

With the introduction of $u_p$, eq. (\ref{eq:stepnl}) becomes 
\begin{equation}
\frac{1}{3} p \frac{d f_0}{d p} (u_2 - u_p) - f_0 \left[u_p+\frac{1}{3} 
p \frac{du_p}{dp} \right] + u_0 f_\infty + Q_0(p) = 0 ,
\label{eq:step1}
\end{equation}

The solution of this equation can be written in the following implicit form:
$$
f_0(p) = f_0^{reacc} (p) + f_0^{inj}(p) =
$$
$$
\int_{p_0}^{p} \frac{d{\bar p}}{{\bar p}} 
\frac{3 \left[u_0 f_\infty ({\bar p}) 
+ Q_0({\bar p})\right]}{u_{\bar p} - u_2} \times
$$
\begin{equation}
\exp\left\{-\int_{\bar p}^p \frac{dp'}{p'} 
\frac{3}{u_{p'} - u_2}\left[u_{p'}+\frac{1}{3}p' \frac{du_{p'}}{d p'}\right]
\right\}.
\label{eq:solut}
\end{equation}
In the case of monochromatic injection with momentum $p_{inj}$ at the shock 
surface, we can write
\begin{equation}
Q_0(p) = \frac{\eta n_{gas,1} u_1}{4\pi p_{inj}^2} \delta(p-p_{inj}),
\end{equation}  
where $n_{gas,1}$ is the gas density immediately upstream ($x=0^-$) and $\eta$ 
parametrizes the fraction of the particles crossing the shock which are 
going to take part in the acceleration process. 

In terms of $R_{sub}$ and $R_{tot}$, introduced above, the density 
immediately upstream is $n_{gas,1}=(\rho_0/m_p)R_{tot}/R_{sub}$.
We can introduce the dimensionless quantity $U(p)=u_p/u_0$ so that
$$
f_0(p) = \frac{3 R_{tot}}{R_{tot} U(p) - 1} 
\int_{p_0}^{p} \frac{d{\bar p}}{{\bar p}} f_\infty ({\bar p})\times 
$$
$$
\exp\left\{-\int_{\bar p}^p \frac{dp'}{p'} 
\frac{3R_{tot} U(p')}{R_{tot} U(p') - 1}\right\} +
$$
$$
\left(\frac{3 R_{sub}}{R_{tot} U(p) - 1}\right) 
\frac{\eta n_{gas,1}}{4\pi p_{inj}^3}\times 
$$
\begin{equation}
\exp \left\{-\int_{p_{inj}}^p 
\frac{dp'}{p'} \frac{3R_{tot}U(p')}{R_{tot} U(p') - 1}\right\}.
\label{eq:laeffe}
\end{equation}
The structure of the fluid upstream of the shock and the corresponding
spectrum of accelerated particles is determined if the velocity field 
$U(p)=u_p/u_0$ is known. The nonlinearity of the problem reflects in the 
fact that $U(p)$ is in turn a function of $f_0$ as it is clear from the 
definition of $u_p$. 
In order to solve the problem we need to write the equations for the 
thermodynamics of the system including the gas, the reaccelerated cosmic 
rays, the cosmic rays accelerated from the thermal pool and the shock itself. 

The velocity, density and thermodynamic properties of the fluid
can be determined by the mass and momentum conservation equations, with the 
inclusion of the pressure of the accelerated particles and of the preexisting
cosmic rays. We write these equations between a point far upstream 
($x=-\infty$), where the fluid velocity is $u_0$ and the density is 
$\rho_0=m n_{gas,0}$, and the point where the fluid upstream velocity is 
$u_p$ (density $\rho_p$).
The index $p$ denotes quantities measured at the point where the
fluid velocity is $u_p$, namely at the point $x_p$ that can be reached
only by particles with momentum $\geq p$.

The mass conservation implies:
\begin{equation}
\rho_0 u_0 = \rho_p u_p.
\label{eq:mass}
\end{equation}
Conservation of momentum reads:
\begin{equation}
\rho_0 u_0^2 + P_{g,0} + P_{CR,0} = \rho_p u_p^2 + P_{g,p} + P_{CR,p},
\label{eq:pressure}
\end{equation}
where $P_{g,0}$ and $P_{g,p}$ are the gas pressures at the points 
$x=-\infty$ and $x=x_p$ respectively, and $P_{CR,p}$ is the pressure
in accelerated particles at the point $x_p$ (we used the symbol $CR$
to mean {\it cosmic rays}, in the sense of {\it accelerated particles}).
The mass flow in the form of accelerated particles has reasonably been 
neglected.

Our basic assumption, similar to that used in \cite{eich84a}, is that
the diffusion is $p$-dependent and more specifically that the diffusion
coefficient $D(p)$ is an increasing function of $p$. Therefore the typical
distance that a particle with momentum $p$ moves away from the shock is
approximately $\Delta x\sim D(p)/u_p$, larger for high energy particles
than for lower energy particles\footnote{For the cases of interest, $D(p)$
increases with $p$ faster than $u_p$ does, therefore $\Delta x$ is a
monotonically increasing function of $p$.}. As a consequence, at each 
given point $x_p$ only particles with momentum larger than $p$ are able 
to affect appreciably the fluid. Strictly speaking the validity of the 
assumption depends on how strongly the diffusion coefficient depends on 
the momentum $p$.

The cosmic ray pressure at $x_p$ is the sum of two terms: one is the pressure
contributed by the adiabatic compression of the cosmic rays at upstream 
infinity, and the second is the pressure of the particles accelerated or 
reaccelerated at the shock (${\tilde P}_{CR}(p)$) and able to reach the 
position $x_p$. Since only particles with momentum larger than $p$ can reach 
the point $x=x_p$, we can write
$$
P_{CR,p} = P_{CR,0} \left(\frac{u_0}{u_p}\right)^{\gamma_{CR}} + 
{\tilde P}_{CR}(p) \simeq
$$
\begin{equation}
\simeq P_{CR,0}(p) \left(\frac{u_0}{u_p}\right)^{\gamma_{CR}} + 
\frac{4\pi}{3} \int_{p}^{p_{max}} dp p^3 v(p) f_0(p),
\label{eq:CR}
\end{equation}
where $v(p)$ is the velocity of particles with momentum $p$, $p_{max}$ 
is the maximum momentum achievable in the specific situation under 
investigation, and $\gamma_{CR}$ is the adiabatic index for the 
accelerated particles. In Eq. \ref{eq:CR} the first term represents
the adiabatic compression of the pressure of the seed particles advected
from upstream infinity, while the second term represents the pressure
in the freshly accelerated particles at the position $x_p$. 

In the following we use $\gamma_{CR}=4/3$ (see \cite{blasi2004} for
a detailed discussion of the reasons for this choice).

The pressure of cosmic rays at upstream infinity is simply
\begin{equation}
P_{CR,0}=\frac{4\pi}{3} \int_{p_{min}}^{p_{max}} dp p^3 v(p) f_\infty(p),
\end{equation}
where $p_{min}$ is some minimum momentum in the spectrum of seed particles.

From eq. (\ref{eq:pressure}) we can see that there is a maximum distance, 
corresponding to the propagation of particles with momentum $p_{max}$ 
such that at larger distances the fluid is unaffected by the accelerated 
particles and $u_p=u_0$.

We will show later that for strongly modified shocks the integral in eq. 
(\ref{eq:CR}) is dominated by the region $p\sim p_{max}$. This improves
even more the validity of our approximation $P_{CR,p}=P_{CR}(>p)$.
This also suggests that different choices for the diffusion coefficient
$D(p)$ may affect the value of $p_{max}$, but at fixed $p_{max}$ the 
spectra of the accelerated particles should not change in a significant way.

Assuming an adiabatic compression of the gas in the upstream region, 
we can write
\begin{equation} 
P_{g,p}=P_{g,0} \left(\frac{\rho_p}{\rho_0}\right)^{\gamma_g}=
P_{g,0} \left(\frac{u_0}{u_p}\right)^{\gamma_g},
\label{eq:Pgas}
\end{equation}
where we used mass conservation, eq. (\ref{eq:mass}). The gas pressure 
far upstream is $P_{g,0}=\rho_0 u_0^2/(\gamma_g M_0^2)$, where $\gamma_g$ 
is the ratio of specific heats for the gas ($\gamma_g=5/3$ for an ideal 
gas) and $M_0$ is the Mach number of the fluid far upstream.

We introduce now a parameter $\xi_{CR}$ that quantifies the relative weight 
of the cosmic ray pressure at upstream infinity compared with the pressure
of the gas at the same location, $\xi_{CR}=P_{CR,0}/P_{g,0}$. Using this
parameter and the definition of the function $U(p)$, the equation for momentum 
conservation becomes

$$
\frac{dU}{dp} \left[ 1 - \frac{\gamma_{CR}}{\gamma_g} \frac{\xi_{CR}}{M_0^2}
U^{-(\gamma_{CR}+1)} - \frac{1}{M_0^2} U^{-(\gamma_g+1)} \right] + 
$$
\begin{equation}
\frac{1}{\rho_0 u_0^2} \frac{d{\tilde P}_{CR}}{dp} = 0.
\end{equation}
Using the definition of ${\tilde P}_{CR}$ and multiplying by $p$, this 
equation becomes
$$
p\frac{dU}{dp} \left[ 1 - \frac{\gamma_{CR}}{\gamma_g} \frac{\xi_{CR}}{M_0^2}
U^{-(\gamma_{CR}+1)} - \frac{1}{M_0^2} U^{-(\gamma_g+1)} \right] = 
$$
\begin{equation}
\frac{4\pi}{3 \rho_0 u_0^2} p^4 v(p) f_0(p),
\label{eq:eqtosolve}
\end{equation}
where $f_0$ depends on $U(p)$ as written in eq. (\ref{eq:laeffe}). Eq. 
(\ref{eq:eqtosolve}) is therefore an integral-differential nonlinear 
equation for $U(p)$. The solution of this equation also provides the 
spectrum of the accelerated particles.
 
The last missing piece is the connection between $R_{sub}$ and $R_{tot}$, the
two compression factors appearing in eq. (\ref{eq:solut}). The compression 
factor at the gas shock around $x=0$ can be written in terms of the Mach number 
$M_1$ of the gas immediately upstream through the well known expression
 
\begin{equation}
R_{sub} = \frac{(\gamma_g+1)M_1^2}{(\gamma_g-1)M_1^2 + 2}.
\end{equation}
On the other hand, if the upstream gas evolution is adiabatic, then the Mach
number at $x=0^-$ can be written in terms of the Mach number of the fluid at
upstream infinity $M_0$ as
$$
M_1^2 = M_0^2 \left( \frac{u_1}{u_0} \right)^{\gamma_g+1} =
M_0^2 \left( \frac{R_{sub}}{R_{tot}} \right)^{\gamma_g+1},
$$
so that from the expression for $R_{sub}$ we obtain
\begin{equation}
R_{tot} = M_0^{\frac{2}{\gamma_g+1}} \left[ 
\frac{(\gamma_g+1)R_{sub}^{\gamma_g} - (\gamma_g-1)R_{sub}^{\gamma_g+1}}{2}
\right]^{\frac{1}{\gamma_g+1}}.
\label{eq:Rsub_Rtot}
\end{equation}

Now that an expression between $R_{sub}$ and $R_{tot}$ has been found, eq.
(\ref{eq:eqtosolve}) basically is an equation for $R_{sub}$, with the boundary 
condition that $U(p_{max})=1$. Finding the value of $R_{sub}$ (and the 
corresponding value for $R_{tot}$) such that $U(p_{max})=1$ also provides 
the whole function $U(p)$ and, through eq. (\ref{eq:solut}), the distribution 
function $f_0(p)$ for the particles resulting from acceleration and 
reacceleration in the nonlinear regime. When the backreaction of the 
accelerated particles is small, the {\it test particle} solution is 
recovered. 
\begin{figure}[thb]
 \begin{center}
  \epsfig{file=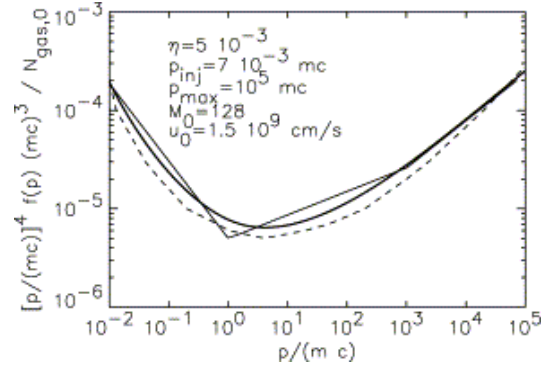,width=7.cm}
  \caption{Example of concave non-linear spectrum obtained with the 
calculation of \cite{blasi02,blasi2004} and described here. The parameters
used are listed in the figure.}
 \end{center}
\end{figure}
\begin{figure}[thb]
 \begin{center}
  \epsfig{file=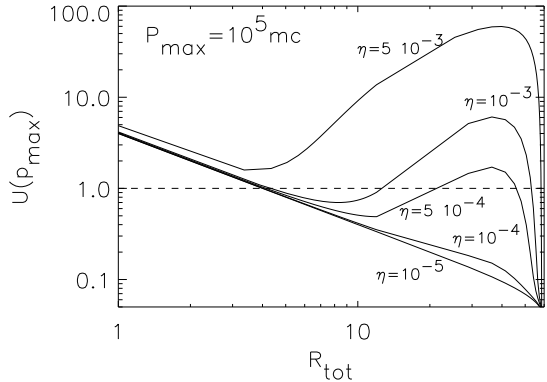 ,width=8.cm}
  \caption{Example of the procedure that leads to multiple solutions. The 
different curves are obtained for the values of the parameter $\eta$ indicated
in the plot.}
 \end{center}
\end{figure}

\begin{figure}[thb]
 \begin{center}
  \epsfig{file=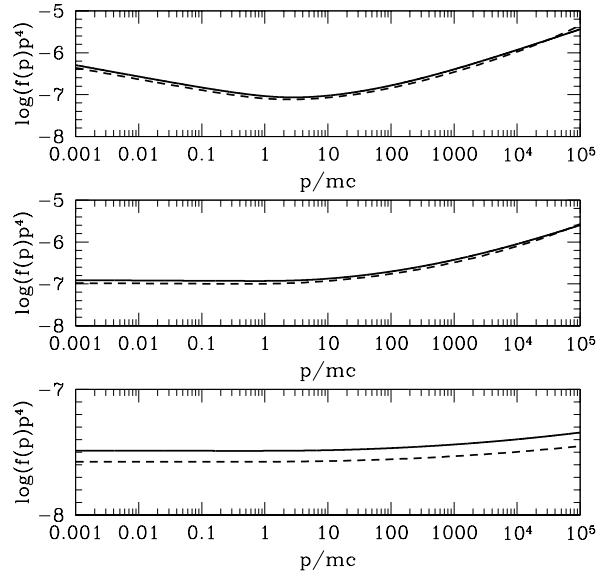,width=8.cm}
  \caption{The figure contains three panels, each one illustrating the
spectrum and the function $U(p)$ for the three solutions obtained using
a specific set of parameters \cite{compmal}. The solid lines indicate the
results obtained with the calculation described here, while the dashed
lines are the results obtained with the method of \cite{malkov1,malkov2}.}
 \end{center}
\end{figure}

\subsection{Non-linear spectra and the problem of multiple solutions}

In Fig. 1 we show an example of the spectrum calculated for parameters 
which are typical of a supernova remnant (solid line), as compared with 
the spectrum estimated according with the simple model of \cite{simple} 
(broken line) and the result of a numerical simulation (dashed line),
also reported in\cite{simple}.

In this calculation no seed particles have been assumed to be present
in the shock environment. The good agreement between the semi-analyical 
approach discussed here and the montecarlo simulations proves that the
semi-analytical approach discussed here is quite effective in describing
the behaviour of cosmic ray modified shock waves as particle accelerators.

However the situation is in general more complex than this: previous 
approaches to the problem of cosmic ray modified shocks had already shown
the appearance of multiple solutions. This was first discussed in 
\cite{dr_v81} in the context of two-fluid models and in \cite{malkov1,malkov2} 
by using a kinetic approach. Multiple solutions are found with the method 
proposed here as well. 
In \cite{blasi2004} it was pointed out how the multiple
solutions appear also in the case of reacceleration of seed particles. 
An example of the phenomenon is illustrated for the case of no seed
particles in Fig. 2, where we plot $U(p_{max})$, bound to be unity for
the physical solutions, as a function of the total compression factor
$R_{tot}$. Here $p_{inj}$, $p_{max}$ and the shock Mach number are all
fixed. The solutions are identified by the points of intersections
of the curves (obtained for different values of $\eta$, as indicated) 
with the horizontal line at $U(p_{max})=1$. 

One can see that for low values of $\eta$ (approximately unmodified
shock) there is only one intersection at $R_{tot}\approx 4$. However,
when $\eta$ is increased the intersections may become three. All the three
solutions are fully acceptable from the point of view of conservation 
laws. For large values of $\eta$ the shock is always strongly modified 
($R_{tot}\gg 4$).
For these cases, the asymptotic shape of the spectrum at large momenta
is well described by the power law $f(p)\sim p^{-7/2}$ (or $N(E)\sim 
E^{-3/2}$). 

The comparison between the method described above and that of 
\cite{malkov1,malkov2} has been discussed in \cite{compmal}. In Fig. 3, 
extracted from \cite{compmal}, we illustrate the spectra and $U(p)$ for 
a case in which three solutions appear (in both approaches). 
The case corresponds to Mach number $M=150$, 
gas temperature at upstream infinity $T=10^4~K$, injection momentum 
$p_{inj}=10^{-3}~mc$ and maximum momentum $p_{max}=10^5~mc$. 
In the calculations of \cite{malkov1,malkov2} a specific
form for the diffusion coefficient as a function of momentum is required.
For reference we adopted a Bohm diffusion coefficient $D(p)\propto p$. 
In Fig. 3, each panel corresponds to one solution. We plot in each panel 
the spectrum $f(p)$ multiplied by $p^4$ (the linear theory would predict
$f(p)p^4 = constant$). The solid lines show the spectra as calculated with 
the approach of \cite{blasi02,blasi2004}, while the dashed lines are the
corresponding spectra as obtained using the calculations of 
\cite{malkov1,malkov2} with Bohm diffusion. The agreement between the
two methods is excellent, despite the fact that the approach presented
here does not require the detailed knowledge of the diffusion coefficient.

\begin{figure}[thb]
 \begin{center}
  \epsfig{file=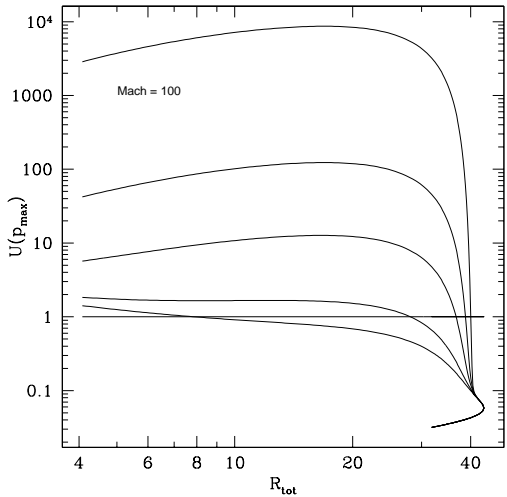 ,width=7.cm}
  \caption{Example of disappearance of multiple solutions with the thermal 
leakage recipe.}
 \end{center}
\end{figure}

\subsection{Are multiple solutions related to the injection problem?}

The question arises of whether the appearance of multiple solutions is
an artifact of our ignorance of the parameter $\eta$, which defines the 
efficiency of the shock in injecting particles from the thermal pool.
Although this is probably not the all story, as confirmed by the fact that 
multiple solutions are present even in the case of reacceleration 
of pre-accelerated particles (in that case $\eta$ is no longer a free 
parameter) \cite{blasi2004}, it is likely 
that injection plays a crucial role. In order to show this, we adopt a simple
physical recipe for the injection of particles at the shock. {\it Real}
shock fronts are not one-dimensional sheets but rather complex surfaces 
with a typical thickness that for collisionless shocks is expected to 
be of the order of the Larmor radius of the thermal protons downstream of the
shock. One should keep in mind that the temperature of the downstream gas
is also affected by the non-linear modification induced by the accelerated
particles, therefore the shock is expected to be thinner when the subshock
is weaker. Our recipe is the following: we assume that the particles which
are injected at the shock are those with momentum $p>\xi p_{th}$, where 
we choose $\xi=3.25$ and $p_{th}$ is the momentum of the thermal particles
in the downstream plasma (we assume that the gas distributions are Maxwellian),
determined as an output of the non-linear calculations from the Rankine-
Hugoniot relations at the subshock. This approximation is sometimes called
{\it thermal leakage} \cite{gieseler}. In physical terms, this makes $\eta$
an output of the calculations rather than a free parameter to be decided 
{\it a priori}. 

In Fig. 4 we plot $U(p_{max})$ calculated as described above, in the 
case in which $\eta$ is evaluated self-consistently from the prescription
of thermal leakage. The different curves are obtained for $p_{max}=
5\times 10^{10}, 10^7, 10^5, 10^3, 10^2~mc$ (from top to bottom) 
for a fixed Mach number $M=100$. One can see that only single intersections 
with the horizontal line $U(p_{max})=1$ are present, namely the multiple
solutions disappear if the shock is allowed to determine its own level 
of efficiency in particle acceleration. This calculation was repeated for
different values of the parameters, but the conclusion was confirmed for
all cases of physical interest \cite{vannoni}. One can also see that large
values of $p_{max}$ typically correspond to more modified shocks, and that
the compression factor can reach large values, far from the test particle
prediction. 

\section{Discussion and General Remarks}

We discussed some aspects of particle acceleration in astrophysical 
collisionless shock waves, and showed that even when the fraction of
particles that participate in the acceleration process is relatively 
small (one in $10^4$ of the particles crossing the shock surface) a large
fraction of the incoming energy can be channelled into few non-thermal
particles. This result, found previously by using several different 
approaches, is of the greatest importance for the physics of cosmic 
rays. Not only the accelerated particles can keep a substantial fraction
of the energy available at the shock, but the spectrum of the accelerated
particles may substantially differ from a power law, showing a concavity
which appears to be the clearest evidence for the appearance of cosmic 
ray modified shocks. Despite the passive role that electrons are likely 
to play in the shock dynamics, the spectrum of accelerated electrons 
is expected to be determined by the (cosmic ray modified) velocity 
profile determined by the accelerated hadrons in the shock vicinity.
A concavity in the spectrum of the radiation generated by relativistic
electrons appears to be one of the possible evidences for shock acceleration 
in the non-linear regime. In the case of supernova remnants, there are
hints that this concavity might have been observed \cite{snr}. 

One of the aspects of particle acceleration that are more poorly understood
is the injection of particles from the thermal pool of particles crossing
the shock. This ignorance reflects in the difficulty of determining the
fraction of particles that takes part in the acceleration process, and we
argued that this might be the reason (or one of the reasons) why calculations
of the non-linear shock structure may show the appearance of multiple
solutions. On the other hand, assuming a simple recipe for the injection 
process (thermal leakage) is shown to result in the existence of only one 
solution. In other words, if the heating and acceleration processes are 
interpreted as two aspects of the same physical phenomenon, there seem to 
be no ambiguities in the way the shock is expected to behave. In this case,
there is no doubt that strongly modified shocks are predicted.

The efficient particle acceleration at strong shocks is also expected to 
result in the reduced heating of the downstream plasma, as compared with 
the heating achieved in the absence of accelerated particles. This effect
should be visible in those cases in which it is possible to measure the 
temperatures of the upstream and downstream fluids separately, for instance
through the X-ray emission of the thermal gases. When the shock is strongly 
modified by the accelerated particles, a large fraction of gas heating is 
due to adiabatic compression in the shock precursor, rather than to shock
heating at the gasous subshock.

In \cite{blasi2004} it was pointed out that if the shock propagates in a 
medium which is populated by seed pre-accelerated particles, the non-linear
modification of the shock can be dominated by such seeds rather than by the
acceleration of fresh particles from the thermal pool. This might be the
case for shocks associated with supernova remnants, which move in the
interstellar medium where the cosmic rays are known to be in rough pressure 
balance with the gas. The spectra of re-accelerated particles for modified
shocks were calculated in \cite{blasi2004} and showed the usual concavity 
that is typical of cosmic ray modified shocks. 

There is an additional aspect of particle acceleration at shock waves that 
has not been discussed so far, namely the generation of a turbulent magnetic
field in the upstream section, due to the streaming instability induced by 
the accelerated particles. The fact that the pressure in the form of 
accelerated particles may reach an appreciable fraction of the kinetic 
pressure at upstream infinity, $\rho_0 u_0^2$, suggests that the magnetic 
field can also be amplified to a turbulent value which may widely exceed 
the background magnetic field, and approach the equipartition level. 
In \cite{bl,bell2004} the process of amplification has been studied 
numerically, and this naive expectation has been confirmed. 
One should however notice that the non-linear effects in particle 
acceleration, discussed in this paper, and in particular the spectral 
modification, are not included self-consistently in the calculations 
of the field amplification in the shock vicinity. 

All these issues are relevant for the investigations of the origin of 
ultra-high energy cosmic rays in many ways: 1) strongly modified shocks
can be very efficient accelerators, so that the energy requirements for
the sources we know might be substantially relaxed; 2) the spectra of
particles accelerated at strongly modified shocks are flatter than those
expected in the linear theory. Flat spectra generate a GZK feature which 
is milder than that due to steep spectra, therefore it may be a less 
severe problem to explain possible excesses of events at the highest energies;
3) magnetic field amplification in the shock vicinity has been invoked in 
the case of SNR's as a possible way to accelerate particles up to the
ankle in these sources \cite{bl,drury,ptuskin}. For other classes of 
sources this may imply that it is easier to reach ultra-high energies
in cases that are currently believed to have too low magnetic fields.
This last point deserved deeper investigation.

\end{document}